\newcounter{myctr}

\documentclass{ws-acs}
\usepackage{url}

\begin{document}

\makeatletter
\def\@biblabel#1{[#1]}
\makeatother

\markboth{R.~A.~Blythe}{Null model for language dynamics}

%
\catchline{}{}{}{}{}
%

\title{NEUTRAL EVOLUTION: A NULL MODEL FOR LANGUAGE DYNAMICS}


\author{\footnotesize RICHARD A.\ BLYTHE}

\address{SUPA, School of Physics and Astronomy, University of Edinburgh\\
Mayfield Road, Edinburgh EH9 3JZ, UK\\
R.A.Blythe@ed.ac.uk}

\maketitle

\begin{history}
\received{(received date)}
\revised{(revised date)}
\end{history}

\begin{abstract}
We review the task of aligning simple models for language dynamics with relevant empirical data, motivated by the fact that this is rarely attempted in practice despite an abundance of abstract models.  We propose that one way to meet this challenge is through the careful construction of null models.  We argue in particular that rejection of a null model must have important consequences for theories about language dynamics if modelling is truly to be worthwhile.  Our main claim is that the stochastic process of neutral evolution (also known as genetic drift or random copying) is a viable null model for language dynamics. We survey empirical evidence in favour and against neutral evolution as a mechanism behind historical language changes, highlighting the theoretical implications in each case.
\end{abstract}


\section{Why model language dynamics?}

There has been a surge in enthusiasm for modelling human social behaviour over the past few years. Language dynamics is particularly popular, both among a community of modellers (who typically have a background in statistical physics \cite{cas09}) and latterly with linguists and psychologists too \cite{cri03,hru09}.  However, there also seems to be some lingering discontent regarding the lack of contact between formal models on the one hand, and empirical data on the other.  

For example, Castellano, Loreto and Fortunato state early on in their excellent review of mathematical models for social dynamics \cite{cas09} that ``there is a striking imbalance between empirical evidence and theoretical modelization, in favor of the latter. \ldots The introduction of a profusion of theoretical models has been justified mainly by vague plausibility arguments, with no direct connection to measurable facts. '' (p593 of \cite{cas09}).  Simply put, the models in question may display rather beautiful physics, but have limited empirical utility due to featuring impoverished treatments of both the behaviour transmitted between agents and of the agents themselves.

Although some resistance to the quantitive modelling of language dynamics has been reported among certain linguists (see \cite{mar09}), it seems nevertheless a reasonable endeavour \emph{as long as} one does not attempt to recreate single instances of historical change with a computational model. For, as Hruschka \textit{et al} observe, ``[g]iven the stochastic nature of language change, trying to predict individual trajectories and particular histories would be a fool's errand'' (p466 of \cite{hru09}).  Instead, these authors advance the use of \emph{null models} as a means to disentangle the various possible factors that may account---in a statistical sense---for general patterns of language change.  It is this idea that I pursue in the present work as a means to proceed towards the laudable and important goal of aligning mathematical models of social dynamics with relevant empirical data.  

The main purpose of a null model is to provide the probability distribution of observable outcomes due to (in principle all) factors other than the one of interest to an experimenter.  Most often a null model incorporates finite sample-size effects and allows one to determine when a difference between two measured values is significant.  However, it is legitimate to include other processes into a null model as long as it is: (i) sufficiently tractable that the required distribution can calculated; (ii) contains sufficiently few free parameters that it is falsifiable; and (iii) intimately related to a specific theoretical hypothesis.  This last point is key, because otherwise rejecting the null model is uninformative.  Further discussion of these three requirements is given in an online appendix to this article.

The remainder of this work is devoted to the proposition that \emph{neutral evolution} (to be defined in the next section) is a viable null model for the dynamics of linguistic variation in a speech community.  In particular, I will show that it is strongly intertwined with the (hotly debated) role of \emph{identity} in language use and change, and to a lesser extent the relevance of meaning to the trajectory of change. Thus, rejection of the null model has important consequences for theories of language change.  I also survey some available empirical evidence for and against neutral evolution as a mechanism for language change. Finally, despite the simplicity of neutral evolution as a mathematical model, I will also demonstrate that there are a number of properties that remain to be established, and thus serve as open problems in the statistical physics of social dynamics.

\section{All roads lead to neutral evolution}
\label{sec:neut}

\emph{Neutral evolution} (also known as \emph{genetic drift} \cite{cro70}, \emph{neutral theory} \cite{hub01} and \emph{random copying} \cite{ben07} in various contexts where it appears) is a process that one is ineluctably drawn to whenever the changes occur due to \emph{replication} acting alone (i.e., without selection) are considered.  Replication sits at the heart of any evolutionary process, and this includes cultural evolutionary processes such as the dynamics of language, wherein a speaker replicates linguistic utterances she has previously heard used to convey a desired meaning.  A model that includes only this process of replication seems like a good starting point in a quest for a null model of language dynamics.

\subsection{The simplest models of replication}
\label{sec:moran}

\begin{figure}[hb]
\centerline{
	\psfig{file=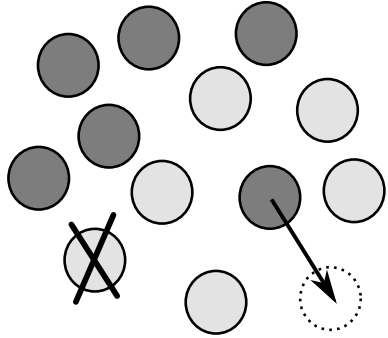,width=5cm}
}
\vspace*{8pt}
\caption{\label{fig:moran} A single update in the Moran model.  One of the tokens (filled circles) is selected at random and duplicated. One existing token is randomly chosen and deleted so that the total number of stored tokens remains constant over time. The illustrated update leads to the variant represented by the darker shading increasing in frequency relative to that represented by the lighter shading.}
\end{figure}

A simple model of replication in a linguistic context can be formulated as follows.  A speaker has a memory of $N$ occasions where a particular meaning ${\cal M}$ was conveyed.  On $n_v$ occasions, a \emph{token} of a particular linguistic structure $v$ was used ($v=1,2,3,\ldots$).  When next called upon to convey the same meaning, she picks one of the $N$ stored tokens with equal probability, and replicates it.  She retains a record of this utterance in her store by replacing one of the existing tokens (again chosen at random) with a token of the variant she just produced. See Figure~\ref{fig:moran}. In a population genetics context, this particular model is known as the Moran model \cite{mor58}.  We will examine various extensions and generalisations, including interactions with other speakers, below.

If $N$ is large, and $x_v$ measures the probability that the speaker uses variant $v$ to convey ${\cal M}$, it is well established that the probability distribution of $x_v$ is governed by the Fokker-Planck (or Kolmogorov, or diffusion) equation
\begin{equation}
\label{ne}
\frac{\partial}{\partial t} P(x_v, t) = \frac{\partial^2}{\partial x_v^2} x_v(1-x_v) P(x_v, t)
\end{equation}
if one unit of time corresponds to $N^2$ token production events.  This is the defining equation of genetic drift \cite{cro70}, a process that has been studied intensively by population geneticists since the 1930s.  One can find extensive discussion of how this important equation is derived and solved in the literature (see e.g., \cite{cro70,ewe04,bly07} and references therein), so I will not repeat this unnecessarily here.

The dynamical rules used above to define this model are completely arbitrary. One would be correct in levelling Castellano \textit{et al}'s criticism of being ``justified mainly by vague plausibility arguments, with no direct connection to measurable facts'' at this model.  However, we launch our defence of neutral evolution as a null model for language dynamics with the observation that \emph{very many models in which replication takes place at a rate that does not depend on the replicator's structure are also described by the diffusion equation (\ref{ne})}.

\begin{figure}[hb]
\centerline{
	\psfig{file=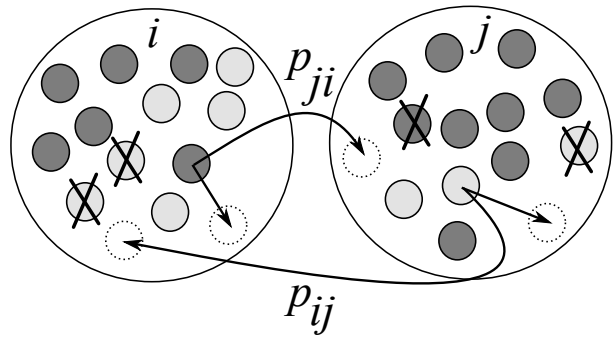,width=8cm}
}
\vspace*{8pt}
\caption{\label{fig:multimoran} A multi-speaker generalisation of the Moran model.  Here, two speakers $i$ and $j$ interact, each sampling a token from their stores. With probability $p_{ij}$ speaker $i$ retains a copy of speaker $j$'s sampled token in his store (displacing an existing token from store $i$).  With probability $p_{ji}$ converse applies.  The standard Moran update (wherein a sampled token is reinserted into a speaker's own store) is always performed for both speakers.}
\end{figure}

Let us first add other speakers to the mix.  This can be achieved by introducing a separate Moran-type population for each speaker.  At any given interaction, two speakers are chosen to interact, and a token is sampled from each store.  Copying a sampled token and replacing it within the same store that the sample was taken from corresponds to a speaker ``listening to herself''---that is, reinforcing her own behaviour.  Additionally, one speaker can place a copy of the other speaker's token in his store: this corresponds to a speaker modifying their behaviour to match more closely that of the interlocutor.  We can assign different probabilities to events of copying tokens between stores---see Figure~\ref{fig:multimoran}. This allows different speakers to be influenced to a greater or lesser extent by different members of the community: as we will discuss in the next section, variation of these probabilities between speakers relates to acts of identity and related phenomena in sociolinguistics.

This multi-speaker generalisation of the Moran model is a version of the \emph{Utterance Selection Model} \cite{bax06}, a mathematical instantiation of the evolutionary model for language change proposed by Croft \cite{cro00}.  Translated into population genetics language, this is genetic drift in a subdivided population \cite{rou04}.  Under certain technical assumptions on the network of interactions between speakers \cite{bly10,bly11}, which seems to amount to the network having the `small-world' property (i.e., a short distance between any pair of speakers, relative to the size of the network, a property believed to be true of real social networks), equation (\ref{ne}) can \emph{still} be used to describe the dynamics of a linguistic variant, albeit under a different definition of time unit, and a weighted average of the usage frequency in place of $x_v$.

Different models for the storage of tokens in memory can be employed, and still give rise to an equation of the form (\ref{ne}). For example, one can relax the rigid one-in, one-out policy that is enforced in the Moran model; alternatively, one can replace more than one token in each update, or even a number drawn from a distribution \cite{ewe04}.  This number can  further depend on the age of the speaker: if the age distribution in the community remains stationary as individuals age, it has  been found that Eq.~(\ref{ne}) continues to describes the dynamics of a variant (again with a suitably defined unit of time) \cite{bax11}.  It seems likely that any combination of the the above effects would serve only to change the characteristic timescale of Eq.~(\ref{ne}).

The essential ingredient of an evolutionary process that leads to an equation of the form (\ref{ne}) is the uniform sampling of stored tokens in producing an utterance. This amounts to an assumption that the different variants $v$ of a linguistic variable are functionally and socially equivalent. In the above, we have described the process in terms of different forms with a common meaning. This may apply to synonyms of the same word, different ways of conveying a grammatical function (like the future tense), or different phonetic realisations of a vowel (such as the `a' in the word `trap'). This is the type of variation typically considered by sociolinguists (see, e.g.,~\cite{lab01}). Fundamentally different dynamics emerge when the sampling is nonuniform.  The simplest way to achieve this is if mutations are possible---that is, once a stored token is chosen, there is some probability that a different token is actually uttered.  This could model the case of attempting to convey a meaning that differs slightly from ${\cal M}$, or the fact that due to articulatory or auditory constraints, a speaker is not able to produce the intended token.  The main effect of mutations on the Moran model is that it supports a stationary state in which multiple variants coexist \cite{cro70,ewe04}.  In all models that are described by (\ref{ne}), every variant bar one eventually goes extinct.  The effect of mutations could (and should) be incorporated into a null model if, for example, there is evidence that variation is being generated as the language evolves.

Fundamentally different behaviour also arises when the probability a variant is used is a nonlinear function of its frequency in the store.  Generically, such nonlinearities can be interpreted as a form of selection, and hence departure from a neutral model---which may or may not include mutation as appropriate to the empirical context---would normally be taken to indicate that selection played a role in shaping the structure of a population. In the context of language dynamics, two selection mechanisms are often advocated. Some take the view that one variant may be inherently more `functional' than another (e.g., because it is more easily articulated)---see e.g.~\cite{net99}.  Croft \cite{cro00} however argues that such functional factors are better represented by mutation processes generating variation: it is then left to \emph{social factors} (such as one variant being associated with a prestigious group of speakers) to propagate some variants at the expense of others. Gotelli and McGill \cite{got06} argues that in population genetics, where reproduction and mutation rates can be independently measured, neutral evolution (as encapsulated by Eq.~(\ref{ne}) or one of the extensions that incorporates mutation) can be used as a null model corresponding to the hypothesis that there is no selection acting.  However, in an ecological context these parameters are less accessible---and in cultural evolutionary applications also certainly cannot be directly observed.  Therefore more work is needed to convince ourselves that neutral evolution can be used as a null model for language dynamics.

\subsection{A more concrete basis in cognition}

In a recent paper, Reali and Griffiths \cite{rea09} (hereafter, RG) forged an important link between explicit models of learning and neutral evolution.  The agents in these models are \emph{Bayesian learners}.  Given some prior beliefs regarding how languages might in principle be structured, and (incomplete) data obtained from interactions with other language users, a Bayesian learner draws a rational inference as to the structure of the language used in his speech community.  An advantage of this approach is that it is closely aligned with empirical research in psychology: Bayesian learning is one of the paradigms that have been used to understand the behaviour of human subjects in a number of laboratory-based experiments (see e.g., the special issue beginning with Ref.~\cite{smi08}).

The crucial step taken by RG was to integrate Bayesian learning into the \emph{iterated learning} paradigm of Kirby and coworkers \cite{smi03}.  Iterated learning generically describes the  situation where an agent's task is to learn a structure that has been learned before.  RG operationalise this with a \emph{diffusion chain}, wherein the first agent is assumed already to have learned a language.  This agent then uses the language, and the second agent in the chain infers its structure from the utterances that are produced.  Once this second agent has reached a certain age, the first agent is removed from the system, and the second agent produces utterances in the presence of a third agent and so on.

As the nomenclature suggests, the learning algorithm in this model exploits Bayes' theorem.  The learner's seeks to determine the frequency $x_v$ that variant $v$ is used to convey meaning ${\cal M}$ by the teacher. If $N$ tokens are heard in total, $n_v$ of which are of variant $v$, the learner constructs the probability distribution for the frequency $x_v$ via Bayes' rule
\begin{equation} 
P(x_v | n_v) \propto P(n_v | x_v) P(x_v)
\end{equation}
in which the constant of proportionality is fixed by summing over all possible $v$.  On the right-hand side of this expression are the \emph{prior} $P(x_v)$ and the likelihood $P(n_v | x_v)$.  The prior encapsulates the learner's prior beliefs about the set of possible language structures.  For example, this could be uniform: anything goes.  Alternatively, it could favour values of $x_v$ close to zero and one, thereby building in a preference for languages in which speakers consistently use one particular variant to convey the meaning, as opposed to switching freely between them.  Another possibility is the opposite preference.  RG allow for all these possibilities by adopting a Dirichlet distribution for the prior:
\begin{equation}
P(x_v) \propto x_v^{\gamma-1} (1-x_v)^{\gamma -1} \;.
\end{equation}
If $\gamma=1$, the prior is uniform; consistency is favoured (disfavoured) for $0 < \gamma<1$ ($\gamma>1$).

The function $P(n_v | x_v)$ is the probability the learner ascribes to the event that $n_v$ tokens of variant $v$ are uttered \emph{assuming} that the speaker's usage frequency is known to be $x_v$.  As in the Utterance Selection Model \cite{bax06}, RG assume a binomial distribution of $N$ trials (tokens) with a success probability $x_v$. Having constructed the posterior distribution $P(x_v)$, RG assume that the learner takes its mean to fix an actual value of $x_v$ to use for his own token productions.  As a consequence of these particular choices, RG find that $x_v$ is governed by a neutral evolutionary dynamics within which mutations between any pair of variants occurs at a constant rate proportional to $\gamma$.  More precisely, one has for large $N$ the Fokker-Planck equation
\begin{equation}
\label{nem}
\frac{\partial}{\partial t} P(x_v, t) = \frac{2\gamma}{k} \frac{\partial}{\partial x_v} (k x_v-1) P(x_v, t) + \frac{\partial^2}{\partial x_v^2} x_v(1-x_v) P(x_v, t)
\end{equation}
where $k$ is the total number of possible linguistic variants conveying the meaning ${\cal M}$ and one unit of time corresponds to $2N$ generations of learners.

This derivation of a neutral evolutionary dynamics from a specific model of cognition provides a firmer basis for the use of neutral evolution as a null model for language dynamics.  It is however true that the direct mapping to a standard neutral evolutionary model depends somewhat crucially on the choice of the prior, the likelihood function, the way the learner selects $x_v$ from the posterior and the fact that one agent learns only from another in a single batch, with the linguistic behaviour remaining fixed after this learning period is over.  As noted in the previous section, a rather large number of changes can be made to the Moran model, whilst still being well-described by the diffusion equation (\ref{ne}).  This is also when mutations are present.  It is therefore plausible that there exist other combinations of Bayesian inference and iterated learning may also lead to a model for the language dynamics of the form (\ref{nem}), although this has not yet been established.

\subsection{Linguistic theories and neutral evolution}

In sociohistorical linguistics, one is often interested in the question of why a language change (e.g., a change in word order) occurred.  A variety of  fundamental mechanisms have been proposed.  For example, the theory of \emph{accommodation} \cite{gil73} describes a process in which interacting speakers become more alike in their linguistic behaviour.  Speakers who frequently interact with one another would be expected to be more similar than those who rarely converse.  A theory that allows for accommodation alone as a mechanism for change may only make reference to different interaction frequencies between pairs of speakers in explaining any observed change process. 

In terms of the multi-speaker Moran process introduced in Section~\ref{sec:moran}, an accommodation-only theory allows for different pairs of speakers to be chosen to interact (have their stores sampled), but the probability that a token is copied from the first to the second speaker must equal that of copying from the second to the first.  The weight assigned to another speaker's utterances depends on the identity of neither speaker nor hearer.

Contrasting theories are those that are based on \emph{prestige} \cite{lab01} or \emph{acts of identity} \cite{lep85}.  Here the identity of speaker and listener may both play in role in determining (a) how much weight a listener gives to a speaker's utterances at the time they are produced and (b) whether, at the time of production, a speaker favours one variant over another due to the identity of speakers who are associated with it.  An example of an act of identity would be a speaker emulating the linguistic behaviour of a socially prestigious group, perhaps acquiring some of this prestige in the process, or at least identifying themselves as aligning with this group as opposed to any other.

The role of identity as a an explanatory factor in language change has been disputed in some quarters. Most notably, Trudgill argues that, in cases of new-dialect formation in emerging societies, the notion of a new national identity played no part in the language changes that took place \cite{tru04,tru08}. Trudgill argues that all such changes took place through the process of accommodation alone, although this is a matter of some debate as the responses to the discussion paper \cite{tru08} demonstrate.

Accommodation- and identity-based theories for language change very nearly map onto neutral and non-neutral models of replication respectively.  This can be understood by considered symmetry relations between different variants and between different speakers. By denying a bias towards or against a particular variant at the time of token retrieval (option (b) above), any process of sampling of stored variants that is not invariant under their relabelling is excluded.  Additionally preventing anything other than interaction frequency affecting token storage demands an invariance under relabelling of speakers as well as variants.  Baxter \textit{et al} \cite{bax06,bax09} observe that the frequency speaker $i$ interacts with speaker $j$ is necessarily equal to the frequency speaker $j$ interacts with speaker $i$.  However, the weight ascribed to speaker $i$'s utterances by speaker $j$ need not equal the converse.  An accommodation-only theory, however, does mandate an equality of these weights. Baxter \textit{et al} \cite{bax08} showed that as a consequence of this symmetry, the social network structure does not affect the choice of time units that yields Eq~(\ref{ne}) (or, if mutations are included, Eq.~(\ref{nem})).  The only demographic factor affecting the unit of time is the overall size of the speech community. The fact that this quantity is directly observable greatly simplifies the application of neutral models to real instances of language change.

The one place where neutral theory and accommodation-based theories diverge is that a linear sampling of the stored tokens is excluded from the former but not necessarily from the latter.  It is therefore possible that one could reject the null hypothesis that a change occurred due to neutral evolution, but this needn't rule out an accommodation-only explanation.  We will return to this point below when we examine the case of the New Zealand English language dialect.

\subsection{Is neutral evolution null enough?}

In the introduction we set out three basic requirements that must be satisfied for a model to enjoy a null status. (These are discussed in more detail in the online appendix).   First, it must be simple enough that the entire distribution of outcomes it predicts can be calculated.  Second, any parameters in the null model must be independently measurable.  Finally, the null model must describe all possible outcomes of a theory that excludes a theoretically interesting factor.  I now argue that neutral evolution broadly satisfies these requirements as a null model for language dynamics, albeit with a few caveats.

\paragraph{Simple enough?} In mathematical terms, the null models of interest here correspond to Eq.~(\ref{ne}), if mutation is thought not to be relevant, Eq.~(\ref{nem}), if it is.  The full time-dependent solutions of these equations for arbitrarily many variants and starting from any initial conditions are available \cite{cro70,gri79,bax07,bly10}.  In principle, therefore, the distribution of any function of variant frequencies can be calculated.

\paragraph{No free parameters?} The question of whether the null model contains free parameters is more vexing. First we have to decide if mutation appears to be present, or not, and if so, at what rate this is occurring.  If there is no evidence for the spontaneous generation of new variants during the period of interest, it would be appropriate to use the mutation-free version of the null model.  It still remains to determine how the unit of time in which Eq.~(\ref{ne}) is defined is related to real time.  In the following section we outline one attempt at this.

When mutation is judged to be relevant, one also needs to measure the mutation rate, ideally independently of the data set which is to be tested against the null model.  One way to do this may be through laboratory experiments, e.g., by estimating the parameter $\gamma$ appearing in Reali and Griffith's model (as was done for example in \cite{rea09cog}).  Alternatively, for certain mutation models, there exist ways to test for departure from neutral evolution that do not require knowledge of either time units or mutation rates \cite{ewe04}.  Such tests may be a useful tool in the analysis of cultural evolutionary processes.

\paragraph{Theoretically interesting?}  The hypothesis that a language change occurred through neutral evolution alone is very nearly theoretically interesting.  There is a close---but not quite exact---mapping between neutral evolutionary processes and those that exclude prestige or identity effects.  There are, however, two corner cases to contend with.  First, the class of neutral models exceeds that allowed by accommodation-only theories, in that it allows for speaker identity to play a role at the time a token is retained in a store.  If possible, one must exclude from the null model those instances where such unwanted effects are present.  The second issue is that accommodation allows for certain nonlinear effects that are not part of a standard neutral evolutionary theory.  If a null model that excludes these nonlinearities were rejected, one would need to check that it would not subsequently be accepted when the nonlinearities are introduced.

\bigskip
In short, despite its imperfections, I regard neutral evolution to be \emph{sufficiently} simple, parameter-free and theoretically interesting to act as a null model for language dynamics.

\section{Neutral evolution versus empirical reality}
\label{sec:apps}

I now examine empirical evidence for and against the null model of neutral evolution as a mechanism for language change, beginning with those cases where the data do not allow the null hypothesis to be rejected.

\subsection{The New Zealand English dialect}

Baxter \textit{et al} \cite{bax09} used neutral evolution as a null model for the formation of the New Zealand English (NZE) language dialect, as documented in the comprehensive work of Gordon \textit{et al} \cite{gor04}, and upon which Trudgill's accommodation-based theory for new-dialect formation was based \cite{tru04}.  Briefly, NZE was formed as a consequence of waves of immigrants from Britain and Ireland arriving in New Zealand in the mid-19th century.  This placed speakers of different regional dialects of English into contact with one another, and over the subsequent two generations, a considerable amount of variation within the speech community was eliminated.

Trudgill \cite{tru04} discusses in detail several linguistic variables (e.g., phonetic realisations of vowel sounds appearing in a class of words).  In each of the cases discussed, one of the variants present in the initial condition is `fixed' in the present-day dialect (i.e., used consistently by all speakers in the community). An estimate of the initial frequency for each successful variant is provided in \cite{tru04}.

Using neutral evolution as to model Trudgill's theory (further constrained to preclude any speaker-based weighting of variants beyond interaction frequency effects), the prediction is that if a variant $v$ is present at a frequency $x_v$ in the initial condition it will fix with probability $x_v$.  To decide whether the observed outcome is consistent with the neutral hypothesis, Baxter et al \cite{bax09} note that with the particular set of frequencies provided by Trudgill \cite{tru04}, the observed outcome (in which two variants initially in the minority fixed), had a likelihood that is about one tenth of the most probable outcome.  That is, if multiple parallel NZE dialects were to form from the same initial condition, ten cases where all the majority variants fixed would be matched by one case where the observed outcome occurred.  This does not appear to be sufficiently rare for the null hypothesis to be rejected.

Another way to quantify the likelihood of the observed outcome under the neutral hypothesis is to apply a type of `exact test' (see e.g.~\cite{sla94}). The idea here is to calculate the cumulative probability of all possible events that have a smaller probability of being realised than the observed event. Again, using the frequency data of \cite{tru04}, the combined probability within the neutral theory of all possible New Zealand English dialects that are less likely than that observed is $74\%$.  This is too large for the null hypothesis of neutrality to be rejected.  Moreover, models that admit nonuniform sampling of stored tokens, but in a way that is invariant under relabelling of variants or speakers, have been found in general to be a \emph{worse} fit to the NZE frequency data than the linear model  \cite{bly09}.  This leads us to believe that neutral evolution is indeed an appropriate null model for new-dialect formation.

\subsection{The Zipf word distribution}

It is well established that the distribution of word frequencies in a language have a Zipf (power-law) distribution. That is, $x_k \sim 1/k^{\sigma}$, where $k$ is the rank of the word ($k=1$ is the most frequent word, $k=2$ the second most frequent etc.), and $\sigma$ is some exponent, usually found to be close to unity \cite{zip49}.  Reali and Griffiths \cite{rea09} have argued that this distribution is to be expected from neutral theory.

To model this situation, RG consider each variant to be a distinct word, and allow in principle an infinite number of them.  That is, when a mutation occurs, a hitherto unused word enters the vocabulary.  This corresponds to the \emph{infinite alleles model} in population genetics \cite{ewe04}.  In this model, words simply continue to be used at approximately their current frequency, subject to finite sample size fluctuations and the occasional introduction of new words.

Recall that within the formulation of the null model presented in Section~\ref{sec:moran}, each variant that exists is taken to have a \emph{single} common meaning ${\cal M}$.  Thus the actual \emph{meaning} of words, and in particular, any relationship between word meanings (e.g., synonymy) plays no role in this model.  That is, two words with the same meaning are considered to be competing in the same space as two words with completely different meanings. RG do not comment on this aspect of their analysis, nor do they refer to a theory that suggests that words with different meanings should (or should not) compete in this way.  Therefore, if the neutral model is found to predict a Zipf distribution of word frequencies (or not), it is somewhat unclear how this finding should be interpreted.

What RG do find can be summarised as follows.  They first take a sample of size $N=33399$, since that corresponds to the size of a widely-used corpus of child-directed speech \cite{ber84}.  On simulating the neutral model with a suitable choice of parameters in the prior, RG find they are able to reproduce the power law with exponent $\sigma = 1.70$ believed to well describe this corpus.  This finding is at odds with that of Fontanari and Perlovsky \cite{fon70} who argue that the variant abundance distribution (when variants are rank ordered) has an exponential tail rather than a power law.  Interestingly, these authors cite a earlier work by Tullo and Hurford \cite{tul03} that proposes neutral evolution as a more appropriate null model for a word distribution than an earlier contender that comprised only random emission of symbols and spaces. This illustrates again the important question of what counts as interesting when designing a null model (see online appendix).

Fontanari and Perlovsky provide one explanation as to why a power-law distribution might be inferred from simulations like those performed by RG, namely finite sample-size effects.  The method used in \cite{fon70} appears to probe the full distribution more efficiently than brute-force methods. However, it is not stated precisely how these finite-size effects would lead to a power law, and whether the samples used to determine empirical word distributions fall into this finite-size regime.  Closer scrutiny of the role of sample size is needed to resolve this conflict.

\subsection{Rate of lexical replacement as a function of frequency}

In a prominent paper, Lieberman \emph{et al} \cite{lie07} established that the rate at which an irregular verb was regularised decreases with its usage frequency.  Whilst linguists have long understood the importance of frequency in terms of linguistic structure (see e.g., \cite{byb01} for a collection of articles on this topic), Lieberman \emph{et al}'s contribution was to state the functional form of a relationship that could be inferred from empirical data.  Specifically, they found that replacement rate of an irregular verb is inversely proportional to the square-root of its frequency.  Reali and Griffiths \cite{rea09} have argued that this functional form is also to be expected from neutral theory.

To apply neutral evolution to this situation, RG took each variant to be a different verb, and asked the question: given that a particular verb is used with frequency $x$, how long does it take to go extinct within the infinite-alleles model?  It was then assumed that when a verb went extinct, it would be replaced by a regular form with the same meaning.  The regularisation rate was then taken to be the reciprocal of the extinction time.  Simulating this process, and plotting the results as a function of the initial frequency, an inverse square-root relationship between initial frequency and replacement rate was found.  However, a mathematical analysis, also presented in \cite{rea09} shows that the inverse square-root law is not exact for neutral evolution (although it is a reasonable fit across a certain range of frequencies).  More stringent statistical tests are needed to determine conclusively whether the data for verb regularisation really is compatible with neutral theory.

What is perhaps more concerning about this formulation is that all other verbs that coexist with the one that is outgoing have the same status.  What is not included in this model is the fact that the regular and irregular form of some verb are competing with each to be used for a common meaning, while the regular forms of two different verbs are not necessarily in competition with each other (at least not in the same way).  Again, it is important to relate the model more directly to linguistic theories for how verbs compete with one another.

\subsection{The trajectory of an innovation}

It is widely recognised among linguists that when a new convention usurps an existing one, the frequency of the innovation follows an `S-curve' trajectory, starting slowly, then gathering pace, before slowing down again as the old form dies out.  In their remarkable work, Reali and Griffiths \cite{rea09}, further claim that neutral evolution exhibits such a phenomenon.

To arrive at this conclusion, RG note an important subtlety.  An equation like (\ref{ne}) describes all possible future trajectories starting from an initial condition.  In particular, the time at which a variant fixes is a random variable.  However, in an instance of observed change, this time is a known quantity.  RG thus argue that one should \emph{condition} on fixation occurring at a particular time.  When this is done, one finds that the \emph{mean} trajectory of the incoming variant, averaged over multiple realisations of the process, follows an S-curve.

However, this does not quite correspond to the empirical situation, in which only one realisation of the process is observed.  Here what is needed is the \emph{probability} of the observed trajectory within the subsensemble of changes that fix at the same time.  As far as I are aware, calculation of this quantity is an outstanding technical challenge.  Until this is achieved, it is hard to know whether conditioning neutral evolution on a given fixation time is sufficient to generate an S-curve with high probability in individual stochastic realisations of the dynamics.

\subsection{The rate of new-dialect formation}

We conclude by returning to the example of New Zealand English.  In addition to the final structure of the new dialect, Baxter \textit{et al} \cite{bax09} also considered the formation time, noting that it was completed in a remarkably short period (approximately $50$ years) given the size of the speech community (some $100,000$ or more speakers).  We recall from Equation (\ref{ne}) that all timescales are proportional to a single characteristic timescale.  In the restricted class of neutral evolutionary models that correspond to the accommodation-only theory advanced by Trudgill \cite{tru04,tru08}, Baxter \textit{et al} found that this characteristic timescale is proportional to the community size, and does not depend on its structure.  The origin of this surprising structure independence is in the symmetry between pairs of speakers that is implied by the accommodation-only theory.  That is, if in any one interaction, each speaker accommodates towards the other by the same degree, only the community size matters.  Community structure may play a role if this symmetry is relaxed and could, for instance, explain the variance of linguistic diversity with community structure noted by Lupyan and Dale \cite{lup10}, though we do not discuss this possibility further here.

Even within the accommodation-only scenario, there remains one important factor that does contribute into the emergent timescale for language change.  This is the time it takes a speaker to forget a stored token.  If the community is large, this time must be short to allow the change to proceed at the observed pace. However, it cannot be arbitrarily short: it certainly can't be shorter than the time between any two utterances.  Experiments on infants \cite{rov95} suggest that any reasonable estimate for this time should in fact exceed two days.  Already, given the size of the speech community, this leads to a characteristic timescale for the formation of the new dialect that vastly exceeds the two generations that was actually observed.  Whilst there remains considerable uncertainty in the appropriate choice for the shorter timescale in this model, it seems likely that the null hypothesis of neutral evolution driving the formation of New Zealand English can be rejected.

\section{Challenges for the future}

In this work I have critically addressed the question of how to apply simple models of language dynamics to relevant empirical data.  As an alternative to the construction of models whose microscopic rules appear to be informed largely by guesswork, I advocate a systematic approach based around a null model which, when rejected, implies a theory of a fundamentally different kind to the corresponding null hypothesis.  I have further proposed that neutral evolution, as defined in Section~\ref{sec:neut}, is an appropriate null model for language dynamics, in that it mostly satisfies the requirements set out in the introduction.

The main selling point is that this model is robustly arrived at from diverse starting points that have one particular feature in common: the only factor affecting language dynamics is the frequency that different variant forms for a particular meaning are used by a speaker's interlocutors.  At the most basic level this corresponds to a \emph{usage-based} theory for language dynamics  in which linguistic structures that emerge are mostly determined by interactions with other language users (see e.g., \cite{tom03}). (These theories contrast with \emph{generative} approaches which typically ascribe a much greater importance to innate aspects of cognition in predetermining the structure of languages).  Another factor in favour of neutral evolution as a null model is that it excludes such complicating factors as the meaning of words or the social status of interlocutors.  Finally, this model is well studied, particularly in the population genetics literature, and many of its predictions are precisely known.

The evidence that some aspects of language dynamics might be accounted for by neutral evolution was reviewed in Section~\ref{sec:apps}.  Here it was found that neutral evolutionary models corresponding to a theory for language change driven purely by accommodation could quantitatively describe the structure of the New Zealand English language dynamics but not the speed at which it was formed.  Although there were free parameters in the model used for this analysis, these conclusions were found to be independent over the range of values they could reasonably take (see \cite{bax09} for full details).  On the other hand, the quantitative support for neutral evolution provided by Reali and Griffiths \cite{rea09}, namely compatibility with word-frequency distributions and lexical replacement rates, involved parameter fitting in both cases.  According to Gotelli and McGill \cite{got06} parameter fitting is problematic because it biases the null hypothesis in favour of being accepted.  Meanwhile, as we noted above in the case of verb regularisation, it is not always clear what the null hypothesis corresponding to neutral evolution is.  This is of crucial importance in understanding the theoretical consequences of rejecting the null model.

From the discussion of Section~\ref{sec:apps} it is apparent that applying a model even as simple as neutral evolution to empirical data can be difficult in practice.  In part this can be due to the nature of available data: it is always difficult to test a theory when data are limited.  This, however, is a state of affairs that suggests effort should be expended on the collection of more relevant data, rather than on the development and study of new models, echoing the sentiment of Castellano \textit{et al} referred to in the introduction.  An important contribution that modellers can make here is to determine precisely \emph{what} data would most effectively yield the deepest insights into the fundamental mechanisms driving language dynamics.

We have, however, also seen that despite the venerability of neutral evolution, a number of technical challenges stand in the way of its being fully understood.  The space of models that has been shown to be characterised at some level by neutral evolution is growing on a daily basis.  As was discussed in Section~\ref{sec:neut}, many modifications can be made to the most simple neutral models whose only effect is to change its single characteristic timescale. What is missing is a precise understanding of what factors lead to a fundamentally different dynamics, and what these factors correspond to in terms of linguistic theory.  The relationship between simple models for language dynamics and Bayesian learning uncovered by Reali and Griffiths looks like particularly fertile ground for future enquiry.  As we have noted, the robustness of neutral evolution as a model for language dynamics when adopting more abstract starting points suggests that a wide range of combinations of Bayesian and iterated learning should ultimately deliver a neutral evolutionary dynamics of some type or other.  Further experimental research, interpreted with the Bayesian learning formalism, would also appear to be useful in constraining parameters that appear in models of language dynamics.

Even within the simplest models, i.e., those described by Equations (\ref{ne}) and (\ref{nem}), some basic properties seem at best poorly appreciated and at worst unknown to us at this time.  For example, the nature of the distribution of variant frequencies within neutral evolution needs to be more widely appreciated.  Perhaps more importantly, Reali and Griffiths emphasise the effects of conditioning the distribution of trajectories generated by neutral evolution on certain observed properties.  This seems to have received very little attention within the modelling community, a fact that prevents certain empirical properties of language change (like the trajectory followed by an innovation) from being satisfactorily analysed.

Closer study of change trajectories is one way in which departure from neutral behaviour may be detected. Much sociolinguistic research is concerned with the manner by which one linguistic variant usurps another, these processes sometimes taking up to several hundred years (as, for example, in the case of the marking of the future tense in Brazilian Portuguese English \cite{pop07}).  Such records of the past are typically unavailable for ecological changes and it may be that there are new ways to detect the presence of selection by examining the time-course of change.  Meanwhile, there are a number of statistical tests to non-neutral diversity patterns in communities sampled at a single point in time (mostly based on Ewens' sampling formula and its relatives \cite{ewe04,eti05}).  To my knowledge, such tests have yet to be applied to cultural evolutionary data.

In short, to achieve the goal of linking models to data, I argue not for more models and certainly not for the cuteness of a model's behaviour being the sole factor in an evaluation of its importance.  Instead what I believe is needed are null models that incorporate all possible factors that are excluded by an interesting, nontrivial and relevant theoretical viewpoint.  For those cases where one wishes to identify social identity or meaning as being significant in a language change, neutral evolution seems to fit the bill.  No doubt there are other null models that are relevant to other key questions about language dynamics, but I think it is fair to say that considerable effort must be expended if we are to draw these out of the vast array of models that have been presented in the statistical mechanics literature so far.

\appendix

\section{(Online only) Desiderata for a null model}

In the main text I provided three basic criteria that should be satisfied by a valid null model:
\begin{itemize}
\item Sufficiently simple that the distribution of observable outcomes can be calculated.
\item Sufficiently few free parameters that the model can be falsified.
\item Sufficiently strong connection to a relevant theory that rejection of the null model is informative.
\end{itemize}
In this online appendix I expand on the reasoning behind these requirements. 

Most scientists are familiar with the concept of a \emph{null hypothesis} in the application of statistical tests to empirical data.  The textbook \cite{ram97} example comprises a series of measurements that can be partitioned into two groups according to some factor of interest to the experimenter.  For example, one could take a bunch of scientists of a certain age, count how many papers each has written, and then divide them into two groups according to whether they drink coffee or not. A statistical test, such as a $t$-test \cite{ram97}, could then be used to ascertain the probability that the null hypothesis, namely that any observed difference in productivity among coffee drinkers compared to their abstinent counterparts is due to finite sample-size effects alone, is compatible with the data.  If this probability is sufficiently small, one concludes that any difference between the two groups is significant, and can be ascribed to coffee drinking accordingly.

Buried deep within this test is the \emph{null model} that is used to obtain the distribution of differences between groups, given that each group is a sample drawn from a single common parent distribution.  The purpose of the null model, therefore, is to generate all possible realisations of the same experiment, so that the experimenter can assess whether the one that was performed is consistent with assumptions of the null model.  In the case of a $t$-test, the central assumption is that each measured data point is an independent sample from a Gaussian parent distribution.

From this simple example, we can extract two important features that a null model should exhibit.  First, it must be possible to determine the probability distribution of experimentally observable outcomes under the assumption that the null hypothesis is correct.  Second, the model should take into account all uninteresting effects.  Then, if one observes an outcome that is sufficiently unlikely within the null model, one can conclude that this is due to an effect that \emph{is} interesting.

It should be apparent that there is a conflict between the simplicity demanded of a model if the entire distribution of outcomes is to be calculated and the complexity needed to take into account every factor that is not of interest to an experimenter.  There are two even more troublesome questions, however.  First, what counts as \emph{interesting}? And second, how much of one's prior beliefs about the system should be incorporated into the null model?

These questions are inter-related, as our (de)caffeinated scientific writers can illustrate.  We are, presumably (but see below), interested only in a \emph{direct} relationship between coffee drinking and a long CV.  It is, however, possible that the reason why one group of scientists drinks a lot of coffee is that they work together in a collaboration that holds its meetings in a cafe.  This might serve to correlate their publication lists, which in turn leads to one of the assumptions of the null model being violated. Thus there is potential for an effect of coffee-drinking on productivity to be erroneously inferred.  There are two ways to handle this problem: either one can make the experiment more like the null model (by trying to minimise any correlations between data points) or one can make the null model more like the experiment (by choosing a test that allows for the existence of such correlations).  Both amount to the same thing, viz, incorporating additional knowledge of the system into the inference process.

Clearly this practice must be scientifically acceptable. Indeed it is desirable that the null model contains empirically plausible processes that nevertheless exclude the postulated explanatory factor.  There is obvious scope for considerable subjectivity here, and entire books have been written on this subject (see, for example, \cite{got96} for a discussion of the application of a range of null models to ecological data).  What is perhaps unacceptable is the practice of including free parameters in a null model, and fitting these to the data set that the model is then applied to.  As pointed out by Gotelli and McGill \cite{got06}, this results in a reduced probability of rejecting the null hypothesis.  Unfortunately, even when one is careful to avoid it, parameter tuning can enter into the application of statistical mechanical models of social dynamics to empirical data in those rare cases where this has been attempted.  To pick an example at random, an analysis of Brazilian election results \cite{ber02} makes the point that an arbitrary rule is no longer required as an explanatory factor when a particular network topology of interactions is adopted, which improves on a previous model.  However, one can equally criticise this analysis on the grounds that the neither the network structure adopted, its size, nor the intermediate time point that was chosen to compare the model with empirical data were parameters fixed by observations other than agreement with a single set of data. The reason why this practice is problematic can be understood from the extreme case of a null model with sufficient flexibility in its parametrisation that it is compatible with any arbitrary data set.  In this case, the null hypothesis corresponds to a theory that is not falsifiable, which is clearly something to be avoided.  One way to help avoid such pitfalls is to be clear about the hypothesis that is being rejected when improvements to models are being made.

This leads us back to the question of what counts as interesting.  This can only be answered with reference to the prevailing theoretical wind.  Again, taking ecology as an example, Gotelli and Graves ``favor a more balanced view that null models describe the assembly of communities, but do not specify all the details of the colonization process'' (\cite{got96}, p4).  To an outsider, this particular boundary between what is allowed into a null model and what is kept out may seem arbitrary but is motivated by a desire to lump together a large class of models, each one a specific instance of a general theory one wishes to reject.   It is this theoretical input that I propose as being of greatest importance in the development of a null model.  If a theory for social dynamics is based on the hypothesis that a particular factor is (ir)relevant in determining how these dynamics play out, a well-designed null model should in principle allow this hypothesis to be tested with appropriate data.  In order to achieve Castellano \emph{et al}'s goal of ``establishing social dynamics as a sound discipline grounded on empirical evidence'' \cite{cas09}, I would argue that engagement with current theories in linguistics, psychology, sociology and elsewhere is a crucial step.  Careful construction of a null model with reference to these theories is a way that this can be done in practice.

\end{document}